\documentstyle[12pt]{article}
\begin{document}

\setcounter{page}{0}
\thispagestyle{empty}

~\vspace{2cm}

\begin{center} {\Large {\bf
Comment on the ``$\theta$-term renormalization in the (2+1)-dimensional
$CP^{N-1}$
model with  $\theta$ term''}} \\
 \vspace{1.5cm}
 {\large
  I.N.Kondrashuk\footnote{e-mail: IKOND@THSUN1.JINR.DUBNA.SU}\\
 \vspace {0.5cm}
 {\em Bogolubov Laboratory of Theoretical Physics,
  Joint Institute for Nuclear Research,
 141980 Dubna (Moscow Region), Russia}}\\
\vspace{0.5cm}
and\\
\vspace{0.5cm}
 {\large
  A.V.Kotikov\footnote{On leave of absence from Particle Physics
    Laboratory, JINR, Dubna, Russia.\\ e-mail:
    KOTIKOV@LAPPHP0.IN2P3.FR; KOTIKOV@SUNSE.JINR.DUBNA.SU}\\
 \vspace {0.5cm}
 {\em Laboratoire de Physique Theorique ENSLAPP\\
LAPP, B.P. 110, F-74941, Annecy-le-Vieux Cedex, France}}
\end{center}

\vspace{4.5cm}\noindent
\begin{center} {\bf Abstract} \end{center}

It is found that the recently published first coefficient of
nonzero $\beta$-function
for the Chern-Simons term in the $1/N$ expansion of the $CP^{N-1}$ model
is untrue numerically.  The correct result is given.  The main
conclusions of the paper \cite{1} are not changed.

PACS numbers: 11.15.Pg, 11.10.Gh

\newpage
\pagestyle{plain}

In a recent paper \cite{1}, S.H. Park investigated the $1/N$ expansion
in the (2+1)-dimensional $CP^{N-1}$ model with a Chern-Simons (or
$\theta$-) term and showed that the $\theta$-term does acquire infinite
radiative corrections
in the first order of $1/N$. We repeated these calculations and found
the complete agreement with these conclusions but a different value of
the $\beta$-function of the $\theta$ charge:
$$ \beta( \theta)~=~ \frac{320}{9 \pi^2}~
\frac{\theta^3}{(1+\theta^2)^2}~
\frac{1}{N} $$

The disagreement between our and S.H. Park's results is in the
calculation of diagrams (5a)-(5e) from \cite{1}.
{\it Firstly}, the diagrams (5d) and (5e) can be represented,
 respectively, as (5b) and (5c) but with the reverse orientation of
arrows in one of two circles. Since
every circle has only two $\overline{n}A_{\mu}n$ vertices containing
momentum and one $\overline{n}\alpha n$ vertex, the contribution of
diagram (5b) is   independent of orientation of arrows and will not
change if arrows are     reversed in one of two circles. Hence, the
contributions from (5b) and (5d), for example, do not cancel each other
like it was proposed in \cite{1} but they are summed. The infinite parts
of four diagrams (5b) - (5e) coincide and equal

$$-\mu^{-2\epsilon}\frac{1}{18 \pi^2}~
\frac{\theta}{1+\theta^2}~ \epsilon^{\mu\rho\nu}
p^{\rho}\frac{1}{\epsilon}\frac{1}{N}.$$

{\it Secondly}, our calculation of the contribution of  diagram (5a)
yields a result which is two times as small as Park's one. We assume
that the reason for this may be a wrong double count of orientation of
arrows (change of orientation of ones does not result in a new diagram).
And {\it in the third place,} our last notation is that for the
correspondence between the Lagrangian and Feynman rules, coefficient of
the $\theta$-term in the Lagrangian must be  two times  as large as
one written by author.  Comparing this Lagrangian with one  in the
author's previous paper \cite{2}, we confirm our assumption.

We calculated the singular parts of the contributions of diagrams
(5a), (5b) and (5d) from \cite{1} by the following way. The leading
(at large $p^2$, where $p$ is the external momentum)
contribution of every diagram, which leads to the renormalization of
$\theta$, has the form
$A \epsilon^{\mu\rho\nu} p^{\rho}/(p^2)^{l\epsilon}$,
where $l$ is the loop number and $A$ is the required coefficient.
After the differentiation with respect to  $p^{\sigma},$ the singular
part of every diagram\footnote{more exactly $kR'$ of diagram but in
our case it coincides with the singular part because there are only
$1/\varepsilon$ terms} does not
depend on momentum $p$ and may be found by Vladimirov's method \cite{3},
where external momentum is put equal to zero (in principe, there is a
necessity of  introducing also some masses to preserve a solution from
infrared singularities but this is not the case).

  The whole sum of the infinite parts of
all diagrams (5a) - (5e) are

$$-\mu^{-2\epsilon}\frac{5}{9 \pi^2}~
\frac{\theta}{1+\theta^2}~ \epsilon^{\mu\rho\nu}
p^{\rho}\frac{1}{\epsilon}\frac{1}{N}.$$

To cancel this infinity we must add to the Lagrangian the
corresponding counterterm, which results in the following expression
for  the  bare charge:

$$ \theta_0 = \mu^{-2\epsilon}\left( \theta +  \frac{80}{9 \pi^2}~
\frac{\theta}{1+\theta^2}~
\frac{1}{\epsilon}\frac{1}{N} \right). $$
 From here we can derive the $\beta$-function that is written above.

To conclude, note that the main result of \cite{1} about the
occurrence
of infinite renormalization of the $\theta$-term in the case of the
$1/N$ expansion does not lose its importance. The function $\beta
(\theta)$ is nonzero and all main conclusions of the paper \cite{1} are
not the subject of a critical review in our comment.

Note only that the results of \cite{1} are in contradiction with the
usual week-coupling expansion where the non-renormalization theorem was
established (see \cite{4}). Technically, the appearance of the nonzero
$\beta$-function in the $1/N$ expansion is quite clear. There is
 $1/N$-resummation of the photon propagator. Another
(half-integer) power of $p^2$ is obtained in the ultraviolet range and
ultraviolet singularities start to appear already in the leading order
of the $1/N$ expansion. Will higher order contributions lead to the
permanent saturation of this effect? It is an open question.

Perhaps, the calculation of the next ($1/N^2$) correction might help
to illuminate this process. However, usually the calculation of higher
order contributions in the framework of the $1/N$ expansion is not a
very simple problem.

The authors are grateful to Dr. S.H. Park for a critical review that
allowed us to avoid the incorrect symmetrical factor for the diagrams
(5b)-(5e) in our calculations.

\end{document}